\documentclass[prc,twocolumn,superscriptaddress,showpacs,amssymb,amsmath,amsfonts,aps]{revtex4}
\usepackage{graphicx}
\usepackage{dcolumn}
\begin{document}
\title{Measurement of the Polarized Structure Function
$\sigma_{LT^\prime}$ for $p(\vec{e},e'p)\pi^o$\\ 
in the $\Delta(1232)$ Resonance Region} 

\newcommand*{\mpaa}{|M_{1+}|^2}
\newcommand*{\mpbb}{|E_{1+}|^2}
\newcommand*{\mpcc}{|S_{1+}|^2}
\newcommand*{\mpdd}{|M_{1-}|^2}
\newcommand*{\mpee}{|E_{0+}|^2}
\newcommand*{\mpff}{|S_{0+}|^2}
\newcommand*{\mpgg}{|S_{1-}|^2}
\newcommand*{\mpa}{M_{1+}}
\newcommand*{\mpb}{E_{1+}}
\newcommand*{\mpc}{S_{1+}}
\newcommand*{\mpd}{M_{1-}}
\newcommand*{\mpe}{E_{0+}}
\newcommand*{\mpf}{S_{0+}}
\newcommand*{\mpg}{S_{1-}}

\newcommand*{\UCONN }{ University of Connecticut, Storrs, Connecticut 06269} 
\affiliation{\UCONN } 

\newcommand*{\VIRGINIA }{ University of Virginia, Charlottesville, Virginia 22901} 
\affiliation{\VIRGINIA } 

\newcommand*{\JLAB }{ Thomas Jefferson National Accelerator Facility, Newport News, Virginia 23606} 
\affiliation{\JLAB } 

\newcommand*{\ASU }{ Arizona State University, Tempe, Arizona 85287-1504} 
\affiliation{\ASU } 

\newcommand*{\SACLAY }{ CEA-Saclay, Service de Physique Nucl\'eaire, F91191 Gif-sur-Yvette, Cedex, France} 
\affiliation{\SACLAY } 

\newcommand*{\UCLA }{ University of California at Los Angeles, Los Angeles, California  90095-1547} 
\affiliation{\UCLA } 

\newcommand*{\CMU }{ Carnegie Mellon University, Pittsburgh, Pennsylvania 15213} 
\affiliation{\CMU } 

\newcommand*{\CUA }{ Catholic University of America, Washington, D.C. 20064} 
\affiliation{\CUA } 

\newcommand*{\CNU }{ Christopher Newport University, Newport News, Virginia 23606} 
\affiliation{\CNU } 

\newcommand*{\DUKE }{ Duke University, Durham, North Carolina 27708-0305} 
\affiliation{\DUKE } 

\newcommand*{\GBEDINBURGH }{ Edinburgh University, Edinburgh EH9 3JZ, United Kingdom} 
\affiliation{\GBEDINBURGH } 

\newcommand*{\FIU }{ Florida International University, Miami, Florida 33199} 
\affiliation{\FIU } 

\newcommand*{\FSU }{ Florida State University, Tallahassee, Florida 32306} 
\affiliation{\FSU } 

\newcommand*{\GWU }{ The George Washington University, Washington, DC 20052} 
\affiliation{\GWU } 

\newcommand*{\GBGLASGOW }{ University of Glasgow, Glasgow G12 8QQ, United Kingdom} 
\affiliation{\GBGLASGOW } 

\newcommand*{\INFNFR }{ INFN, Laboratori Nazionali di Frascati, Frascati, Italy} 
\affiliation{\INFNFR } 

\newcommand*{\INFNGE }{ INFN, Sezione di Genova, 16146 Genova, Italy} 
\affiliation{\INFNGE } 

\newcommand*{\ORSAY }{ Institut de Physique Nucleaire ORSAY, Orsay, France} 
\affiliation{\ORSAY } 

\newcommand*{\BONN }{ Institute f\"{u}r Strahlen und Kernphysik, Universit\"{a}t Bonn, Germany} 
\affiliation{\BONN } 

\newcommand*{\ITEP }{ Institute of Theoretical and Experimental Physics, Moscow, 117259, Russia} 
\affiliation{\ITEP } 

\newcommand*{\JMU }{ James Madison University, Harrisonburg, Virginia 22807} 
\affiliation{\JMU } 

\newcommand*{\KYUNGPOOK }{ Kungpook National University, Taegu 702-701, South Korea} 
\affiliation{\KYUNGPOOK } 

\newcommand*{\MIT }{ Massachusetts Institute of Technology, Cambridge, Massachusetts  02139-4307} 
\affiliation{\MIT } 

\newcommand*{\UMASS }{ University of Massachusetts, Amherst, Massachusetts  01003} 
\affiliation{\UMASS } 

\newcommand*{\UNH }{ University of New Hampshire, Durham, New Hampshire 03824-3568} 
\affiliation{\UNH } 

\newcommand*{\NSU }{ Norfolk State University, Norfolk, Virginia 23504} 
\affiliation{\NSU } 

\newcommand*{\OHIOU }{ Ohio University, Athens, Ohio  45701} 
\affiliation{\OHIOU } 

\newcommand*{\ODU }{ Old Dominion University, Norfolk, Virginia 23529} 
\affiliation{\ODU } 

\newcommand*{\PITT }{ University of Pittsburgh, Pittsburgh, Pennsylvania 15260} 
\affiliation{\PITT } 

\newcommand*{\ROMA }{ Universita' di ROMA III, 00146 Roma, Italy} 
\affiliation{\ROMA } 

\newcommand*{\RPI }{ Rensselaer Polytechnic Institute, Troy, New York 12180-3590} 
\affiliation{\RPI } 

\newcommand*{\RICE }{ Rice University, Houston, Texas 77005-1892} 
\affiliation{\RICE } 

\newcommand*{\URICH }{ University of Richmond, Richmond, Virginia 23173} 
\affiliation{\URICH } 

\newcommand*{\SCAROLINA }{ University of South Carolina, Columbia, South Carolina 29208} 
\affiliation{\SCAROLINA } 

\newcommand*{\UTEP }{ University of Texas at El Paso, El Paso, Texas 79968} 
\affiliation{\UTEP } 

\newcommand*{\UNIONC }{ Union College, Schenectady, NY 12308} 
\affiliation{\UNIONC } 

\newcommand*{\VT }{ Virginia Polytechnic Institute and State University, Blacksburg, Virginia   24061-0435} 
\affiliation{\VT } 

\newcommand*{\WM }{ College of William and Mary, Williamsburg, Virginia 23187-8795} 
\affiliation{\WM } 

\newcommand*{\YEREVAN }{ Yerevan Physics Institute, 375036 Yerevan, Armenia} 
\affiliation{\YEREVAN } 

\newcommand*{\NOWNCATU }{ North Carolina Agricultural and Technical State University, Greensboro, NC 27411}

\newcommand*{\NOWGBGLASGOW }{ University of Glasgow, Glasgow G12 8QQ, United Kingdom}

\newcommand*{\NOWJLAB }{ Thomas Jefferson National Accelerator Facility, Newport News, Virginia 23606}

\newcommand*{\NOWSCAROLINA }{ University of South Carolina, Columbia, South Carolina 29208}

\newcommand*{\NOWFIU }{ Florida International University, Miami, Florida 33199}

\newcommand*{\NOWINFNFR }{ INFN, Laboratori Nazionali di Frascati, Frascati, Italy}

\newcommand*{\NOWOHIOU }{ Ohio University, Athens, Ohio  45701}

\newcommand*{\NOWCMU }{ Carnegie Mellon University, Pittsburgh, Pennsylvania 15213}

\newcommand*{\NOWINDSTRA }{ Systems Planning and Analysis, Alexandria, Virginia 22311}

\newcommand*{\NOWASU }{ Arizona State University, Tempe, Arizona 85287-1504}

\newcommand*{\NOWCISCO }{ Cisco, Washington, DC 20052}

\newcommand*{\NOWUK }{ Kentucky, LEXINGTON, KENTUCKY 40506}

\newcommand*{\NOWSACLAY }{ CEA-Saclay, Service de Physique Nucl\'eaire, F91191 Gif-sur-Yvette, Cedex, France}

\newcommand*{\NOWRPI }{ Rensselaer Polytechnic Institute, Troy, New York 12180-3590}

\newcommand*{\NOWUNCW }{ North Carolina}

\newcommand*{\NOWHAMPTON }{ Hampton University, Hampton, VA 23668}

\newcommand*{\NOWTulane }{ Tulane University, New Orleans, Lousiana  70118}

\newcommand*{\NOWKYUNGPOOK }{ Kungpook National University, Taegu 702-701, South Korea}

\newcommand*{\NOWCUA }{ Catholic University of America, Washington, D.C. 20064}

\newcommand*{\NOWGEORGETOWN }{ Georgetown University, Washington, DC 20057}

\newcommand*{\NOWJMU }{ James Madison University, Harrisonburg, Virginia 22807}

\newcommand*{\NOWURICH }{ University of Richmond, Richmond, Virginia 23173}

\newcommand*{\NOWCALTECH }{ California Institute of Technology, Pasadena, California 91125}

\newcommand*{\NOWMOSCOW }{ Moscow State University, General Nuclear Physics Institute, 119899 Moscow, Russia}

\newcommand*{\NOWVIRGINIA }{ University of Virginia, Charlottesville, Virginia 22901}

\newcommand*{\NOWYEREVAN }{ Yerevan Physics Institute, 375036 Yerevan, Armenia}

\newcommand*{\NOWRICE }{ Rice University, Houston, Texas 77005-1892}

\newcommand*{\NOWINFNGE }{ INFN, Sezione di Genova, 16146 Genova, Italy}

\newcommand*{\NOWBATES }{ MIT-Bates Linear Accelerator Center, Middleton, MA 01949}

\newcommand*{\NOWODU }{ Old Dominion University, Norfolk, Virginia 23529}

\newcommand*{\NOWVSU }{ Virginia State University, Petersburg,Virginia 23806}

\newcommand*{\NOWORST }{ Oregon State University, Corvallis, Oregon 97331-6507}

\newcommand*{\NOWMIT }{ Massachusetts Institute of Technology, Cambridge, Massachusetts  02139-4307}

\newcommand*{\NOWCNU }{ Christopher Newport University, Newport News, Virginia 23606}

\newcommand*{\NOWGWU }{ The George Washington University, Washington, DC 20052}

  
\author{K.~Joo}
     \affiliation{\UCONN}
     \affiliation{\VIRGINIA}
     \affiliation{\JLAB}
\author{L.C.~Smith}
     \affiliation{\VIRGINIA}
\author{V.D.~Burkert}
     \affiliation{\JLAB}
\author{R.~Minehart}
     \affiliation{\VIRGINIA}
\author{G.~Adams}
     \affiliation{\RPI}
\author{P.~Ambrozewicz}
     \affiliation{\FIU}
\author{E.~Anciant}
     \affiliation{\SACLAY}
\author{M.~Anghinolfi}
     \affiliation{\INFNGE}
\author{B.~Asavapibhop}
     \affiliation{\UMASS}
\author{G.~Audit}
     \affiliation{\SACLAY}
% ...

\collaboration{The CLAS Collaboration}

\noaffiliation
\begin{abstract}
{The polarized longitudinal-transverse structure function $\sigma_{LT^\prime}$ 
has been measured in the $\Delta(1232)$ resonance region 
at $Q^2=0.40$ and $0.65$~GeV$^2$. Data for the $p(\vec e,e'p)\pi^o$ reaction 
were taken at Jefferson Lab with the CEBAF Large Acceptance Spectrometer (CLAS) using
longitudinally polarized electrons at an energy of 1.515 GeV.  For the first 
time a complete angular distribution was measured, permitting the separation
of different non-resonant amplitudes using a partial wave analysis.  
Comparison with previous beam asymmetry measurements 
at MAMI indicate a deviation from the predicted $Q^2$ dependence of 
$\sigma_{LT^{\prime}}$ using recent phenomenological models.}
\end{abstract}
\maketitle

The $\gamma^*p\rightarrow\Delta^+(1232)$ transition has long served as a 
benchmark for testing nucleon models.  In the $SU(6)$ symmetric quark model, this strong 
magnetic dipole excitation is described as originating from a single quark 
spin flip.  Residual
spin-dependent and tensor-type interactions between the quarks are needed
to explain the $N-\Delta$ mass difference and the small quadrupole 
transition strength observed in partial wave analyses of experimental
pion electroproduction data~\cite{fro99,mer01,joo02}. Understanding the origin of these 
residual interactions and their role in resonance formation and decay 
is a fundamental challenge for modern QCD-inspired hadronic models.  

In particular, the dynamical effects of the pion cloud are predicted to strongly
modify the electromagnetic couplings at sufficiently low $Q^2$.  
Chiral-quark and bag models that incorporate pion 
couplings~\cite{ber88,wal97,sil00,amo00} generally describe the $\Delta(1232)$ 
photocoupling multipoles better than a purely quark/gluon framework~\cite{isg82,cap90}.
Recent dynamical models derived from effective chiral Lagrangians explicitly
treat pion multiple scattering~\cite{sat01,kam01} and predict strong modifications 
to both resonant and non-resonant amplitudes. The important role of the pion
cloud in electromagnetic interactions has been emphasized recently in
heavy baryon chiral perturbation theory~\cite{gel99} and
unquenched lattice QCD~\cite{ale02}. 

Unfortunately, cross section measurements alone do not provide sufficient
information to separate the 
$\Delta(1232)$ excitation reaction mechanisms from 
non-resonant backgrounds and the tails of higher-mass resonances. 
Single spin polarization observables, on the other hand, are directly sensitive to the
interference between resonant and non-resonant processes and together 
with precise cross sections can provide powerful constraints to models.  

In this Rapid Communciation we report new measurements of the longitudinal-transverse
polarized structure function $\sigma_{LT^\prime}$ obtained 
in the $\Delta(1232)$ resonance region using the $p(\vec e,e'p)\pi^o$ reaction.  
Recent measurements of polarization 
observables~\cite{war98,pos01,bar02,kunz03} and unpolarized cross 
sections~\cite{mer01,spar03} for $Q^2<0.2$~GeV$^2$ show disagreement with some dynamical 
models near the $\Delta(1232)$ peak.  However, so far only narrow angular 
and kinematic ranges have been studied, yielding few clues as to the origin
of the discrepancy.  The present experiment was performed at four-momentum transfers $Q^2=0.40$ and 
$0.65$ GeV$^2$ and covers a range of invariant mass $W= 1.1-1.3$ GeV 
with full angular coverage in $\,\cos\,\theta^*_{\pi}$ and
$\phi^*_{\pi}$ in the $p\pi^0$ center-of-mass (c.m.).

The data were taken at the Thomas Jefferson National Accelerator Facility (Jefferson Lab)
using a 1.515 GeV, 100\% duty-cycle beam of longitudinally 
polarized electrons incident on liquid hydrogen target. 
The electron polarization was determined by frequent 
M{\o}ller polarimeter measurements to be 
$0.69\pm0.009$(stat.)$\pm0.013$(syst.). Scattered electrons and protons were detected in
the CLAS spectrometer~\cite{mec03}. Electron triggers were enabled
through a hardware coincidence of the gas Cerenkov
counters and the lead-scintillator electromagnetic calorimeters.  Protons
were identified using momentum reconstruction in the tracking system
and time of flight from the target to the scintillators. Software 
fiducial cuts were used to exclude regions of non-uniform detector response.
Kinematic corrections were applied to compensate for drift chamber
misalignments. The $p\pi^0$ final state was identified by requiring the missing neutral
to have a mass squared between $-0.01$ and $0.05$ GeV$^2$.
Background from elastic Bethe-Heitler radiation was suppressed to 
below 1\% using a combination of cuts on missing mass and 
$\phi^*_{\pi}$ near $\phi^*_{\pi}=0^o$. Target window backgrounds 
were suppressed with cuts on the reconstructed $e'p$ target vertex.  

In the one-photon-exchange approximation, the electroproduction
cross section factorizes as follows:
\begin{equation}
\frac{d\,^5\sigma}{dE_{e'} d\Omega_{e'} d\Omega^*_{\pi}} = \Gamma_v\,\frac{d\,^2\sigma^h}{d\Omega^*_{\pi}},
\end{equation}
where $\Gamma_v$ is the virtual photon flux and $d\,^2\sigma^h$ is the differential cross
section for $\gamma^* p \rightarrow p \pi^0$ with electron beam helicity ($h=\pm1$). 
For an unpolarized target, $d\,^2\sigma^h$ depends 
on the transverse ($\epsilon$) and longitudinal ($\epsilon_L$)  polarization of the virtual photon 
through five structure functions: $\sigma_T,\sigma_L$ and their interference 
terms $\sigma_{TT}$, $\sigma_{LT}$ and $\sigma_{LT^\prime}$:
\begin{eqnarray}
\frac{d\,^2\sigma^h}{d\Omega^*_{\pi}} &=& \frac{p^*_{\pi}}{k_{\gamma}^*} (\sigma_{0} +
h\sqrt{2\epsilon_L(1-\epsilon)}\,\sigma_{LT^\prime}\,\sin\,\theta^*_{\pi}\,\sin\,\phi^*_{\pi}),  \nonumber 
\\
\sigma_{0} &=& \sigma_T+\epsilon_L\sigma_L+\epsilon\,\sigma_{TT}\,\sin^2\theta^*_{\pi}\,\cos\,2\phi^*_{\pi} \nonumber \\
~&+&
\sqrt{2\epsilon_L(1+\epsilon)}\,\sigma_{LT}\,\sin\,\theta^*_{\pi}\,\cos\,\phi^*_{\pi},
\label{eq:str}
\end{eqnarray}
where ($p_{\pi}^*,\theta^*_{\pi},\phi^*_{\pi}$) 
are the $\pi^o$ c.m. momentum, polar and azimuthal angles, $\epsilon = (1+2|\vec{q}\,|^2\,\tan^2(\theta_e/2)/Q^2)^{-1}$, $\epsilon_L=(Q^2/|k^*|^2)\epsilon$, and $k_{\gamma}^*$ and  $|k^*|$
are the virtual photon c.m. momentum 
and equivalent energy.

The structure functions $\sigma_{LT}$ and $\sigma_{LT^\prime}$
determine the real and imaginary parts of bilinear products between
longitudinal and transverse amplitudes:  
\begin{eqnarray}
\sigma_{LT}: Re(L^*T) &=& Re(L)Re(T) + Im(L)Im(T) \\
\sigma_{LT^{\prime}}: Im(L^*T) &=& Re(L)Im(T) - Im(L)Re(T).
\end{eqnarray}
Detection of a weak non-resonant background underlying the peak of the 
$\Delta(1232)$ can be enhanced through its interference 
in $\sigma_{LT^{\prime}}$ with the strong transverse magnetic 
multipole $Im(M_{1+})$.  Sensitivity to real backgrounds is 
suppressed in $\sigma_{LT}$ due to the vanishing of $Re(M_{1+})$ at the resonance pole. 

Extraction of $\sigma_{LT^\prime}$ was made through a measurement of the 
electron beam asymmetry $A_{LT^\prime}$: 
\begin{eqnarray}
A_{LT^\prime} &=&\frac{d\,^2\sigma^+ - d\,^2\sigma^-}{d\,^2\sigma^+ +
d\,^2\sigma^-}  \\ 
&=&
\frac{\sqrt{2\epsilon_L(1-\epsilon)}\,\sigma_{LT^\prime}\,\sin\,\theta^*_{\pi}\,\sin\,\phi^*_{\pi}}{\sigma_{0}}.
\label{eq:altp}
\end{eqnarray}
$A_{LT^\prime}$ 
was obtained by dividing the measured asymmetry $A_m$ by the
magnitude of the electron beam polarization $P_e$:
\begin{eqnarray}
A_{LT^\prime} &=& \frac{A_m}{P_e}  \\
A_m &=& \frac{N_\pi\,^+ - N_\pi\,^-}
{N_\pi\,^+ + N_\pi\,^-}, 
\label{eq:altp_m}
\end{eqnarray}
where $N_{\pi}^{\pm}$ is the number of $\pi^0$ events per incident electron for each
electron beam helicity state.  $A_{LT^\prime}$ was determined 
for individual bins of $(Q^2,W,\cos\theta_{\pi}^*,\phi_{\pi}^*)$.
Normalization factors cancel in Eq.~6, and since acceptance 
studies showed no significant helicity or bin size dependence, 
acceptance factors canceled in $A_m$ as well.  This leaves $A_m$ largely 
free from systematic errors. Radiative corrections were 
applied for each bin using the program 
recently developed by A. Afanasev {\it et al.} for exclusive pion 
electro-production~\cite{aku02}. Corrections were 
also applied to compensate for cross section variations over the width
of each bin. 
The corrected $A_{LT^\prime}$ was multiplied by the 
unpolarized cross section $\sigma_{0}$.
A parameterization
of $\sigma_{0}$ was used, which was obtained from the SAID PR01 solution~\cite{said01} 
fitted to previously measured CLAS data and world data. The structure function 
$\sigma_{LT^\prime}$ was then extracted using Eq.~\ref{eq:altp} by fitting 
the $\phi_{\pi}^*$ distributions. 
Systematic errors for $\sigma_{LT^\prime}$ were dominated by 
uncertainties in determination of the electron beam polarization 
and the parameterized unpolarized cross 
section $\sigma_{0}$. The systematic error for $A_m$ is 
negligible in comparison. Quadratic addition of the individual
contributions yields a total relative systematic error of $< 6\%$.

\begin{figure}[h]
\includegraphics[scale=0.45]{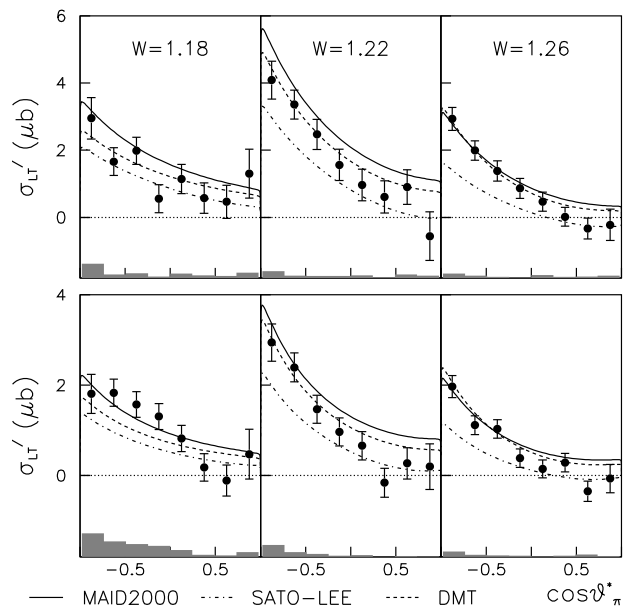}
\caption{CLAS measurement ($\bullet$) of $\sigma_{LT^\prime}$
versus $\cos\theta^*_\pi$ extracted at $Q^2$= 0.40~GeV$^2$ (top) and 
$Q^2$=0.65~GeV$^2$ (bottom).  Curves show model predictions. Shaded bars show systematic errors.}
\label{fig:sig1}
\end{figure}

Fig.~\ref{fig:sig1} shows $\sigma_{LT^\prime}$ extracted at $Q^2$=0.40~GeV$^2$
and $Q^2$=0.65~GeV$^2$, where the $\cos\,\theta_{\pi}^*$ dependence is 
plotted for $W$ bins of 1.18, 1.22, and 1.26~GeV. The measured angular
distributions show a strong backward peaking for $W$ bins around the $\Delta(1232)$
mass.  The curves show predictions
from recent models \cite{sat01,kam99,dre99} which use different methods 
to satisfy unitarity in the $\pi^0 p$ final state.  These models, which are 
fitted to previous photo- and unpolarized electro-production data, include backgrounds 
arising from Born diagrams and $\it{t}$-channel vector meson exchange. 
The Sato-Lee~\cite{sat01} and Dubna-Mainz-Taipei~\cite{kam99} (DMT) models use an 
off-shell $\pi N$ reaction theory to calculate unitarity corrections,
while the more phenomenological MAID2000 model~\cite{dre99} incorporates
$\pi N$ phases directly into the background amplitudes. While the models
describe the data qualitatively, none of the 
calculations is able to describe both the overall magnitude and the slope
of the measured c.m. angular distributions consistently.

\begin{figure}
\includegraphics[scale=0.46]{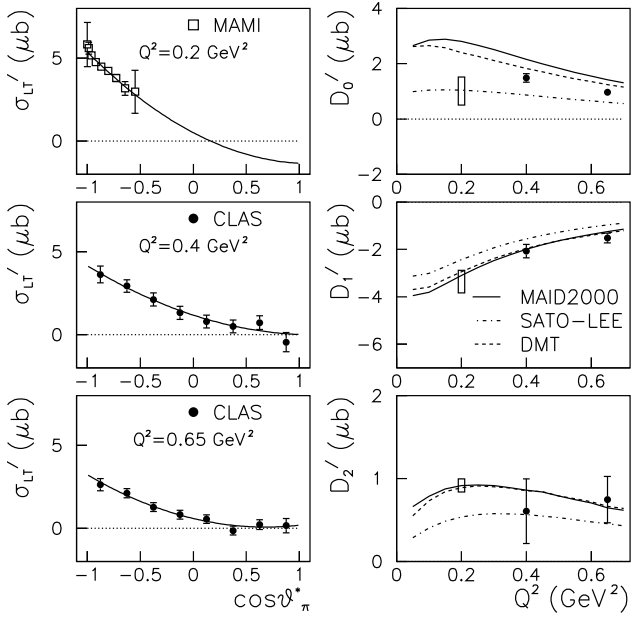}
\caption{LEFT: Fits to $\sigma_{LT^\prime}$ angular distributions 
measured by CLAS (middle,bottom) and MAMI (top) 
at $W=1232$~GeV using Eq.(9).  See text for details.
RIGHT: $Q^2$ dependence of Legendre moments of $\sigma_{LT^\prime}$. 
Curves show model predictions. Data points are the 
present CLAS measurement.  Vertical bars at $Q^2$=0.2~GeV$^2$ show moments
obtained from model constrained fits to MAMI data~\cite{bar02}.}
\label{fig:q2}
\end{figure}

A more quantitative comparison was made through fitting the
extracted $\sigma_{LT^\prime}$ angular distributions using the Legendre 
expansion:
\begin{eqnarray}
\sigma_{LT^\prime} & = & D_0^\prime + D_1^\prime P_1(\cos\theta^*_\pi)
+  D_2^\prime P_2(\cos\theta^*_\pi),
\label{eq:str_legendre} 
\end{eqnarray} 
where $P_l(\cos\theta^*_\pi)$ is the $l^{th}$-order Legendre polynomial
and $D_l^\prime$ is the corresponding Legendre moment. 
Each moment can be decomposed into interference terms involving the leading-order 
magnetic $(M_{l_\pi\pm})$, electric $(E_{l_\pi\pm})$, and scalar
$(S_{l_\pi\pm})$ multipoles: 
\begin{eqnarray}
D_0^\prime &=& -{Im}((M_{1-} - M_{1+} +3E_{1+})^*S_{0+} \nonumber \\
 &+& E_{0+}^*(S_{1-}-2S_{1+}) + ...) \\
D_1^\prime  &=& -6{Im}((M_{1-} - M_{1+} + E_{1+})^*S_{1+} \nonumber \\
 &+& E_{1+}^*S_{1-} + ...) \\
D_2^\prime & = & -12{Im}((M_{2-} -  E_{2-})^*S_{1+} \nonumber \\
 &+& 2E_{1^+}^*S_{2^-} + ...),
\end{eqnarray}
where $l_\pi$ is the $\pi^0 p$ angular momentum whose coupling with the nucleon spin 
is indicated by $\pm$.

Fig.~\ref{fig:q2} shows typical fits to $\sigma_{LT^\prime}$ angular distributions
near the peak of the $\Delta(1232)$ resonance (left), while the $Q^2$ dependence 
of the extracted Legendre moments is compared to model predictions (right).  The largest
disagreement with models clearly occurs for $D_0^\prime$, which is dominated
by interference terms involving $s-$wave $\pi N$ multipoles.  The CLAS data 
also require $D_2^\prime\ne 0$. The fitted $D_2^\prime$ 
strength has the same sign and overall magnitude as the model
predictions, although we cannot differentiate between the models due to large 
statistical uncertainties.  No evidence for $\it{d}-$waves was observed in 
our measurement of $\sigma_{LT}$~\cite{joo02}.

We also compare our fit results with a recent MAMI measurement~\cite{bar02} of the 
beam asymmetry $A_{LT^{\prime}}$ at $Q^2=0.2$~GeV$^2$.  The published MAMI 
angular distribution was converted to $\sigma_{LT^\prime}$ using Eq.~\ref{eq:altp} and 
MAID2000 for the unpolarized cross section $\sigma_0$.  Since the MAMI data do not have
sufficient angular coverage to determine $D_2^\prime$, the fit was performed by
constraining $D_2^\prime$ relative to $D_1^\prime$ using MAID2000.  With $D_2^\prime$
fixed, the remaining Legendre moments estimated from the MAMI data can be compared to
the $Q^2$ trend of the CLAS data (Fig.~\ref{fig:q2}, right). Both data sets suggest
an anomalous behavior for $D_0^\prime$ with respect to the models.  However, a recent BATES 
measurement~\cite{kunz03} of $\sigma_{LT^{\prime}}$ at $Q^2=0.127$~GeV$^2$ 
and $\theta^*_\pi=129^\circ$ found good agreement with MAID2000 and DMT, although
no angular distributions were reported.

\begin{figure}[h]
\includegraphics[scale=0.45]{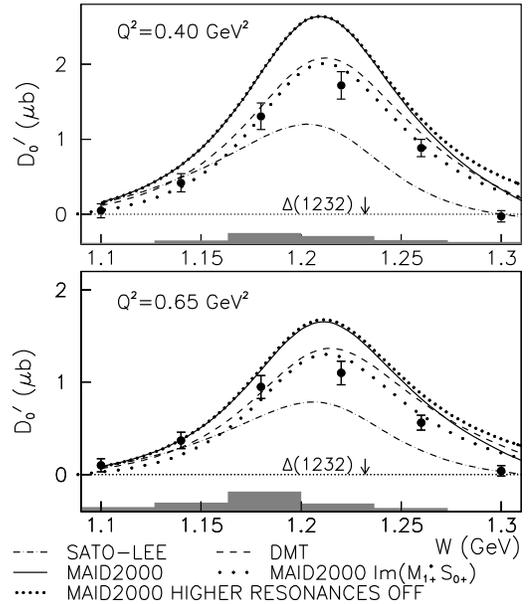}
\caption{CLAS measurement ($\bullet$) of Legendre moment $D_0^\prime$ vs. $W$(GeV). 
Curves show recent model calculations that include contributions from multipoles 
up to angular momentum $l_\pi=5$. Shaded bars show systematic errors.}
\label{fig:d0}
\end{figure}

\begin{figure}[t]
\includegraphics[scale=0.45]{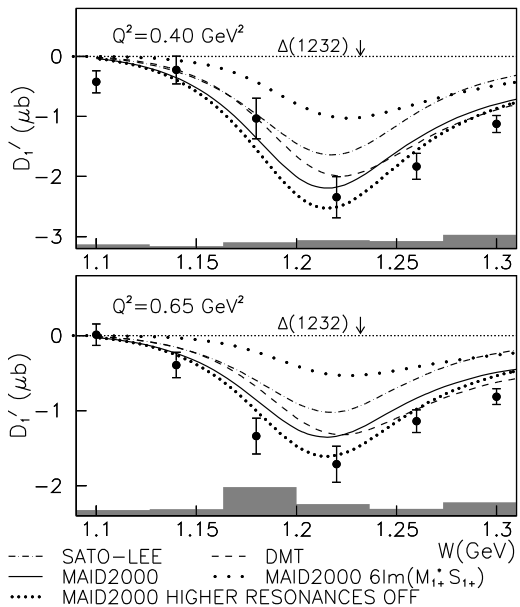}
\caption{CLAS measurement ($\bullet$) of Legendre moment $D_1^\prime$ vs. $W$(GeV). 
See Fig. 3 for details.}
\label{fig:d1}
\end{figure}

Fig.~\ref{fig:d0} and Fig.~\ref{fig:d1} show the $W$ dependence of the 
fitted Legendre moments, $D_0^\prime$ and $D_1^\prime$, respectively.
Both moments show strong resonant behavior, suggesting dominance of
interference terms involving the multipoles of the $\Delta(1232)$. 
 Our measurement of
$D_0^\prime$ is substantially below the predictions of MAID2000, and 
in closer agreement with the DMT dynamical model at $W=1.18$~GeV, while the
Sato-Lee prediction is smaller still.  For increasing $W$, our data fall
below the DMT curve, while none of the models 
describes the $W$ dependence well.  Note that contributions of higher resonances 
to $D_0^\prime$ are negligible except near $W$=1.30 GeV.  
Figure~\ref{fig:d1} shows the fit results for $D_1^\prime$.  Here our
comparison with models shows some $Q^2$ dependence.  Better agreement
with the dynamical models occurs below the $\Delta(1232)$ at 
$Q^2=0.4$~GeV$^2$, while at $Q^2=0.65$~GeV$^2$ all of the $W$ points are 
systematically larger than the predictions.   

The large differences between the model predictions for 
$D_0^\prime$ arise from the term $Im(M_{1+}^*S_{0+})$,
which produces 70-75\% of the total strength in MAID2000. 
In contrast, $D_1^\prime$ is more sensitive to higher resonances, 
which contribute 15-20\% in MAID2000 (coming mainly from $Im(M_{1-}^*S_{1+}))$, 
while $Im(M_{1+}^*S_{1+})$ accounts for $\approx~40\%$ of the total strength. 
The $S_{0+}$ multipole is an important background affecting the extraction of 
the $\gamma^* p\rightarrow \Delta(1232)$ $\it{C2}$ Coulomb quadrupole 
transition, and is sensitive to choices of $\pi NN$ 
coupling and contributions from final state $\pi N$ rescattering~\cite{dre99}.
Unfortunately, a simple rescaling of the $S_{0+}$ strength, as suggested
in Ref.~\cite{bar02}, is not sufficient to account for the inferred 
$Q^2$ dependence of $D_0^\prime$.

In summary, complete angular distributions for the polarized structure 
function $\sigma_{LT^\prime}$ were measured for the first time, using
the  $p(\vec{e},e'p)\pi^o$ reaction.  In accordance with measurements
at lower $Q^2$~\cite{war98,pos01,bar02,kunz03}, evidence for significant non-resonant 
background in the $\Delta(1232)$ region is seen.   A departure from the predicted $Q^2$ 
dependence of various effective Lagrangian based models 
is seen at the $\Delta(1232)$ peak when the CLAS data are compared to the MAMI data at 
$Q^2=0.2$ GeV$^2$. Examination of
the Legendre moments $D_0^\prime$ and $D_1^\prime$ shows the discrepancies are
largest for $D_0^\prime$.  CLAS measurements in the $Q^2$ range of 0.1-0.4~GeV$^2$ 
and also for $W > 1.3$~GeV are 
currently being analyzed to provide more information on the form factors of
the underlying multipoles.

We acknowledge the efforts of the staff of the Accelerator and Physics Divisions at 
Jefferson Lab in their support of this experiment.  This work was supported in
part by the U.S. Department 
of Energy and National Science Foundation, the Emmy Noether Grant from the 
Deutsche Forschungsgemeinschaft, the French Commissariat a l'Energie 
Atomique, the Italian Istituto Nazionale di Fisica Nucleare, and the Korea Research 
Foundation. The Southeastern Universities Research Association (SURA) 
operates the Thomas Jefferson Accelerator Facility for the United States Department 
of Energy under contract DE-AC05-84ER40150.

\end{document}